# Laser Etch Enabled Active Embedded Microfluidic Cooling in β-$Ga_2O_3$

Shonkho Shuvro, Roopa Jayaramaiah, Rangarajan Muralidharan, Digbijoy N. Nath, and Prosenjit Sen

*Centre for Nano Science and Engineering, Indian Institute of Science, Bengaluru 560012, India*

E-mail: prosenjits@iisc.ac.in

We demonstrate active embedded microfluidic cooling in β-$Ga_2O_3$. We employ a cost-effective infra-red laser etch setup to achieve controlled etching of micro-channels in 500 µm thick β-$Ga_2O_3$ substrate. The micro-channels are about 210 µm deep and 340 µm wide. Resistive heating is used as proof-of-concept. At a water flow rate of 50 ml/min, a 50% reduction in surface temperature from ~140°C to ~ 72°C is achieved for 3.5 W of input power. The experimental observations are backed by thermal simulation. This work is expected to lead to a new paradigm in thermal management in emerging β-$Ga_2O_3$ devices.

Recently, β-$Ga_2O_3$ has drawn a lot of attention of the materials and device community because of its superior properties, including its ultrawide bandgap (~ 4.9eV)[1)] resulting in high critical electric field of 8 MV/cm[2)], and high Baliga's Figure of Merit[3)], which makes it promising for beyond- Silicon Carbide (SiC) and beyond- Gallium Nitride (GaN) power electronics. However, its exceptionally poor thermal conductivity (~11W/mK along [100], ~ 27W/mK along [010] & ~17W/mK along [001]) [4)] is a major Achilles' heel for $Ga_2O_3$ which could prevent one from utilizing it to its full potential. Thus, thermal management of $Ga_2O_3$ power devices is a key area of research and is critical towards enabling a technology based on this material system which can pragmatically rival other wide band gap and ultrawide band gap devices.

Although several papers have shown thermal management techniques for $Ga_2O_3$ power devices, they tend to be usually simulation-based studies.[5,6,7)] A common approach for heat management in $Ga_2O_3$ power devices is passive cooling which is essentially integration of high thermal conductivity materials with the chip's hotspot site. By reducing thermal resistance, it prevents the accumulation of heat.[8,9)] Another common and promising approach is to enhance the ability of the chip to transmit heat by achieving effective heat transport with external work. Microfluidic Cooling and Thermo-Electric Cooling (TEC) cooling are two examples of this type of active cooling.[10,11,12,13)] Using convection heat flow to extract heat from confined areas, microfluidic cooling can significantly lower power device temperatures.[14)] Using microchannels that are etched into the device substrates, embedded cooling is an effective method that brings coolant into direct contact with the device substrate. The primary benefit of embedded cooling is that it nearly overcomes all the thermal resistances found in remote cooling.[15)] Such embedded cooling has been demonstrated in GaN-based High Electron Mobility Transistors[16)] with unprecedented improvement in device performance. However, microfluidic channel based embedded cooling has not yet been demonstrated for $Ga_2O_3$. In this work, we report for the first time, embedded microfluidic cooling in β-$Ga_2O_3$ substrate. We also report on a versatile, simple, and cost-effective approach to create such microfluidic channels in gallium oxide using an IR laser which provides many advantages over conventional plasma-based dry etch methods.

We used commercially procured bulk β-$Ga_2O_3$ substrate of thickness 500 μm for this work. A 2-inch wafer was diced, and one such diced pieces was taken for this study. For etching microfluidic channels on the backside of the β-$Ga_2O_3$ substrate, plasma-based dry etching in Reactive Ion Etch (RIE) chamber is a highly challenging task, given that the channels need to be a couple of hundred microns deep and typical dry etch rates are only a few tens of nm/min to a few hundreds of nm/min leading to significant process challenges.[17,18,19] Here, we have used near Infrared (IR) laser with ~30 μm spot size to etch deep microfluidic channels on the backside of β-$Ga_2O_3$ substrate. The laser power and number of cycles are carefully optimized to get the required etch depth. The laser-based etching of β-$Ga_2O_3$ happens due to multi-photon absorption explained elsewhere.[20] Using this technique, we achieved etch rates as high as ~ 30 μm/min. Laser etching holds great promises such as: (i) Creating micro-channels on the backside of the substrate where coolant can be circulated for effective heat management of the high-power devices and circuits. Excessive heat generation is a serious bottleneck that prevents one from realizing the true potential Gallium Oxide devices (ii) Creating via holes in substrate such which is then filled with metal for creating a Ground plane to the devices on the top of the substrate (iii) Thinning of the substrate for heat management. A thinner substrate allows for better heat flow from the active device to the heat sink below (iv) Devices can be isolated using laser ablation.

The fabrication process flow of resistive heating of Gallium Oxide with embedded microfluidic channels is shown in Fig. 1. It begins with the laser etching of β-$Ga_2O_3$, followed by resistive contact formation on the top surface of the gallium oxide substrate using Aluminum metal which are 5 mm in length, 300nm in thickness deposited using e-beam evaporation. The laser etching process is found to result in a tapered profile of the etched microfluidic channels as shown in the Scanning Electron Microscope (SEM) image in Fig. 2 (a) which are about 211 μm wide and 340 μm deep. Next, the microfluidic channels are encapsulated with Silicon wafer through epoxy bonding. Fig. 2(b) shows an optical microscope image of the Al-metal resistive network. This metal resistive network is carefully aligned such that it is exactly on the top of the microfluidic channel to have efficient heat transfer. The experimental setup is shown in Fig. 3. We have used GW Instek GDP-2303S which was calibrated before the experiment as the source unit which can supply up to 30V/3A. The temperature gradient distribution on the top surface is measured using an infrared camera (Fluke Ti400). To inject the coolant into the channels, we have a syringe pump (NE-8000) giving a controlled flow rate. The outcoming coolant is collected in a small container.

Using the experimental setup, the surface temperature of β-$Ga_2O_3$ substrate with and without microfluidic cooling is analyzed. The input power is estimated by the product of input current (which is sourced) and the measured voltage drop. Fig. 4. shows the surface temperature vs. the input power for various flow rates. Here, we have shown three different flow rates which are 10ml/min, 30ml/min, and 50ml/min. Without cooling, the surface temperature reaches 139.4°C corresponding to an input power of 3.5 W. Such a temperature is quite realistic a number for high power devices as reported in various studies.[21,22,23)] For simulations we have used COMSOL Multiphysics software. Here we have considered incompressible, laminar steady-state flow. We have used symmetrical boundary conditions to reduce the computational complexity. The temperature curve is measured experimentally, with the findings being compared to those computed numerically, to ensure that the finite-element model is accurate. The physical parameters and boundary conditions of the finite-element models are given in Table I. Initially when the input power is small the temperature difference for all the flow rates is the same. As the temperature rises the benefits of microfluidic cooling are observed. At the flow rate of 50ml/min, the temperature drops to 72.6°C which is almost 50% reduction in temperature. This is because as the flow rate increases, the convective heat capacity increases and therefore the temperature drops. Fig. 5. (a) shows the simulation results without cooling. Fig. 5. (b) shows the simulation results with cooling (Flow Rate: 50ml/min).

In conclusion, active embedded microfluidic cooling was investigated which shows promising results in terms of reducing the temperature of Gallium Oxide. Through the fabrication of micro-channels in Gallium Oxide Substrate, the coolant is directly introduced into it to offer extremely effective cooling. At the flow rate of 50 ml/min, the surface temperature drops from 139.4°C to 72.6°C which is nearly 50% drop. Also, simulation is consistent with the experimental data. Therefore, to enhance the performance of Gallium Oxide devices, integrated microfluidic cooling holds a great promise as an effective thermal management strategy.

**Acknowledgments**
This work is funded by National Mission on Power Electronics (NAMPET III), Ministry of Electronics & IT (MeitY). We acknowledge the funding from MeitY, DST and MHRD for supporting the facilities. The authors would like to thank the staff at the National Nanofabrication Center (NNFC), Micro and Nano Characterization Facility (MNCF) and Packaging Lab of CeNSE. We also thank Prof. V. R. Supradeepa, Aniruddhan Gowrisankar, Anirudh Venugopalrao, Taha Saquib, Jyotiranjan Sahoo, Rutvik Lathia, Bheema Reddy, and Abhigyan Goswami of CeNSE, IISc, Bangalore.

## References

1) M. Orita, H. Ohta, M. Hirano, and H. Hosono, Appl. Phys. Lett., **77**, 4166 (2000).

2) S. I. Stepanov,V.I. Nikolaev, V.E. Bougrov and A.E. Romanov, Rev. Adv. Mater. Sci., **44**, 63 (2016).

3) Y. Yao., Raveena G. and Jaewoo K., J. Vac. Sci. Technol. B, Nanotechnol. Microelectron., Mater., Process., Meas., Phenomena, **35**, 03D113 (2017).

4) Z. Guo, A. Verma, X. Wu, F. Sun, A. Hickman, T. Masui, A. Kuramata, M. Higashiwaki, D. Jena, and T. Luo, Appl. Phys. Lett. **106**, Art. no.111909 (2015).

5) B. Chatterjee, K. Zeng, C. D. Nordquist, U. Singisetti, and S. Choi, IEEE Trans. Compon., Packag., Manuf. Technol., **9**, 2352, (2019).

6) C. Yuan, Y. Zhanf, R. Montgomery, S. Kim, J. Shi, A. Mauze, T. Itoh, J. S. Speck, and S. Graham, Appl. Phys. Lett, **127**, 15402 (2020).

7) Y. Mao, B. Meng, Z. Qin, B. Gao and C. Yuan, IEEE Trans. on Elect. Devices (Early Access), 1(2023).

8)T. Matsumae, Y. Kurashima, H. Umezawa, K. Tanaka, T. Ito, H. Watanabe and H. Takagi, Appl. Phys. Lett **116**, 141602 (2020).

9) Z. Cheng, V. D. Wheeler, T. bai, J. Shi, M. J. Tadjer, T. Feygelson, k. D. Hobart, M. S. Goorsky and S. Graham, Appl. Phys. Lett, **116**, 141602 (2020).

10) C. Kim, M.Jeong, S. Kim, S. Oh, S. Lee, Y. Joo, H. Shen, H. Lee, J. Yoon and Y. Joo, Proc. IEEE 70th Electron. Compon. Technol. Conf. (ECTC), 2020, p.2242.

11) M. Manno, B. Yang, S. Khanna, P. McCluskey, and A. Bar-Cohen, IEEE Trans. Compon., Packag., Manuf. Technol., **5**, 1775 (2015).

12) Z. He, Y. Yan, and Z. Zhang, Energy, **216**, 119223 (2021).

13) C. Green, P. Kotte, X. Han, C. Woodrum, T. Sarvey, P. Asrar, X. Zhang, Y. Joshi, A. Federov, S. Sitaraman and M. Bakir, J. Electron Packag., **137**, 040802 (2015).

14) D. B. Tuckerman and R. F. W. Pease, IEEE Electron Device Lett., **2**,126 (1981).

15) A. Bar-Cohen, J. J. Maurer, and D. H. Altman, J. Electron. Packag., **141**, 040803 (2019).

16)R.v. Erp, R. Soleimanzadeh, L. Nela, G. Kampitsis and E. Matioli, Nature, **585**, 211 (2020).

17) H. Okumura and T. Tanaka, Jpn. J. Appl. Phys, **58**, 120902 (2019).

19) R. Khaana, K. Bevlin, D. Geerpuram, J. Yang, F. Ren, and S. Pearton, *Gallium Oxide Technology, Devices and Applications*, ed. Stephen Pearton, Fan Ren, and Micheal Mastro (Elsevier, Amsterdam, 2019) p.263.

19) L. Zhang, A. Verma, H. Xing, and D. Jena, Jpn. J. Appl. Phys, **56**, 030304 (2017).

20) V. Nathan and A. H. Guenther, J. Opt. Soc. Am., **2**, 294(1985).

21) J. W. Pomeroy, C. Middleton, M. Singh, S. Dalcanale, M. J. Uren, K. Sasaki, A. Kuramata, S. Yamakoshi, M. Higashiwak, and M. Kuball, IEEE Electron Device Lett., **40**, 189 (2019).

22) M. H. Wong, Y. Moriwaka, K. Sasaki, A. Kuramata, S.Yamakoshi and M. Higashiwaki, Appl. Phys. Lett, **109**, 193503 (2016).

23) H. Zhou, K. Maize, J. Noh, A. Shakouri and P. D. Ye, ACS Omega, **2**, 7723 (2017).

24) J. C. Mendes, M. Leiher, C. Li, Materials, **15**, 415 (2022)

25) N. A. B-Abdun, Z. M. Razlan, S. A. Bakar, C. H. Voon, W. K. Wan, I. Zunaidi, M K Albzeirat, and N. Z. Noriman, IOP Conf. Ser.: Mater. Sci. Eng., **557**, 012051 (2018)

## Figure Captions

**Fig. 1.** Fabrication Process flow of $Ga_2O_3$ embedded microfluidic structure. (a) Microfluidic channels using laser etching (b) Metal deposition for resistive networks (c) Epoxy bonding with Si to encapsulate the channels.

**Fig. 2.** (a) SEM Cross-Section Image of Microfluidic Channels etched in Gallium Oxide (b) Resistive Network on Gallium Oxide (Microfluidic channels are visible) with Silicon bonded.

**Fig. 3.** Experimental Setup to inject coolant into the substrate while studying the temperature rise.

**Fig. 4.** $Ga_2O_3$ Surface Temperature vs. Input Power for Various Flow Rates. (a) Simulation data is compared with experimental data.

**Fig. 5.** Simulation Result (a) Without Cooling (b) Cooling with Flow Rate of: 50ml/min

Table 1: Simulation Parameters

| Thermal Conductivity | | Boundary Condition | |
|---|---|---|---|
| $Ga_2O_3$ | 11W/mK along [100], 27W/mK along W/mK along [010] & 17W/mK along [001] 4) | Input Flow Rate | 0-50ml/min |
| Si | 150W/mK 24) | Coolant Temperature | 27°C |
| | | Ambient Temperature | 27°C |
| DI Water | 0.57W/mK 25) | Heat Flux | 1.28 - 44.8W/mm$^2$ |
| | | Trench Depth | 340.38μm |
| | | Trench Opening | 211.58μm |
| | | Si Thickness | 0.5mm |
| | | $Ga_2O_3$ Thickness | 0.5mm |

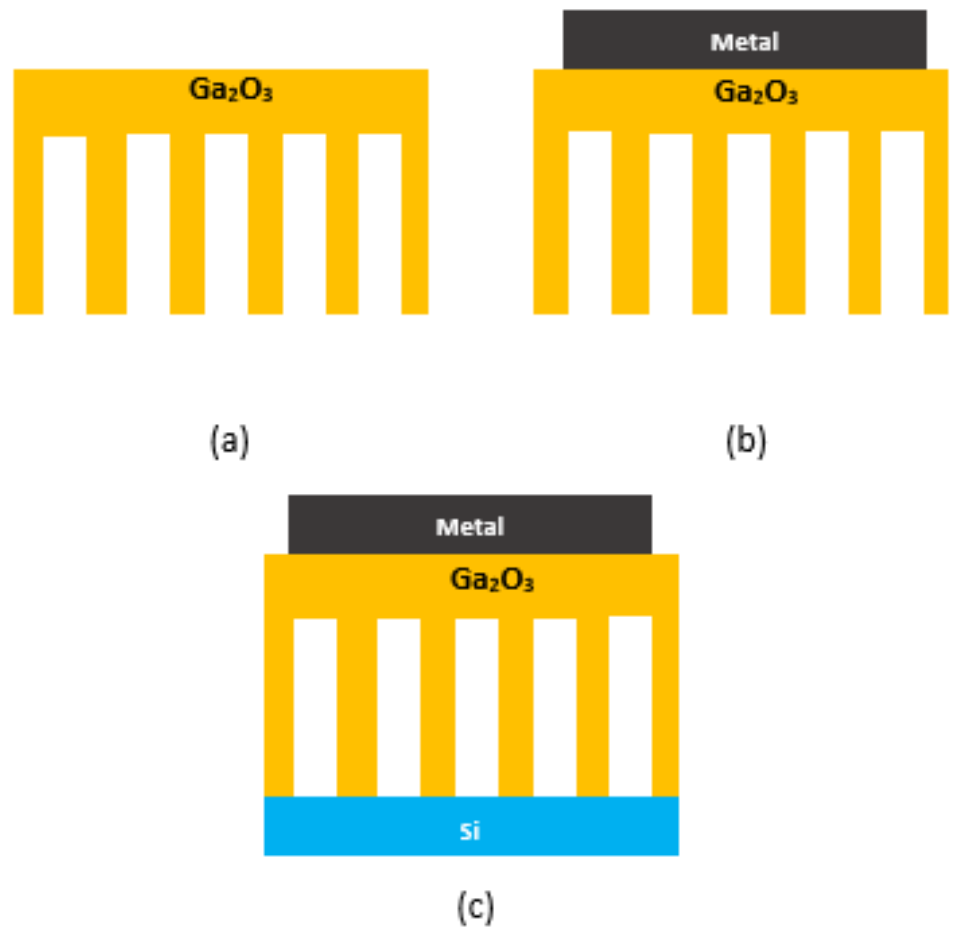

Fig. 1.

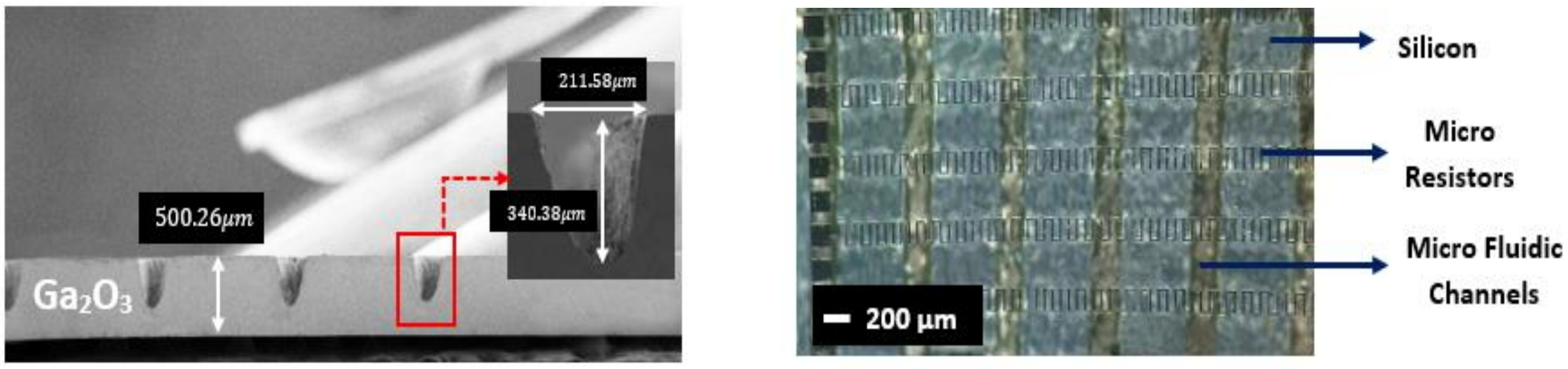

(a) (b)

Fig. 2

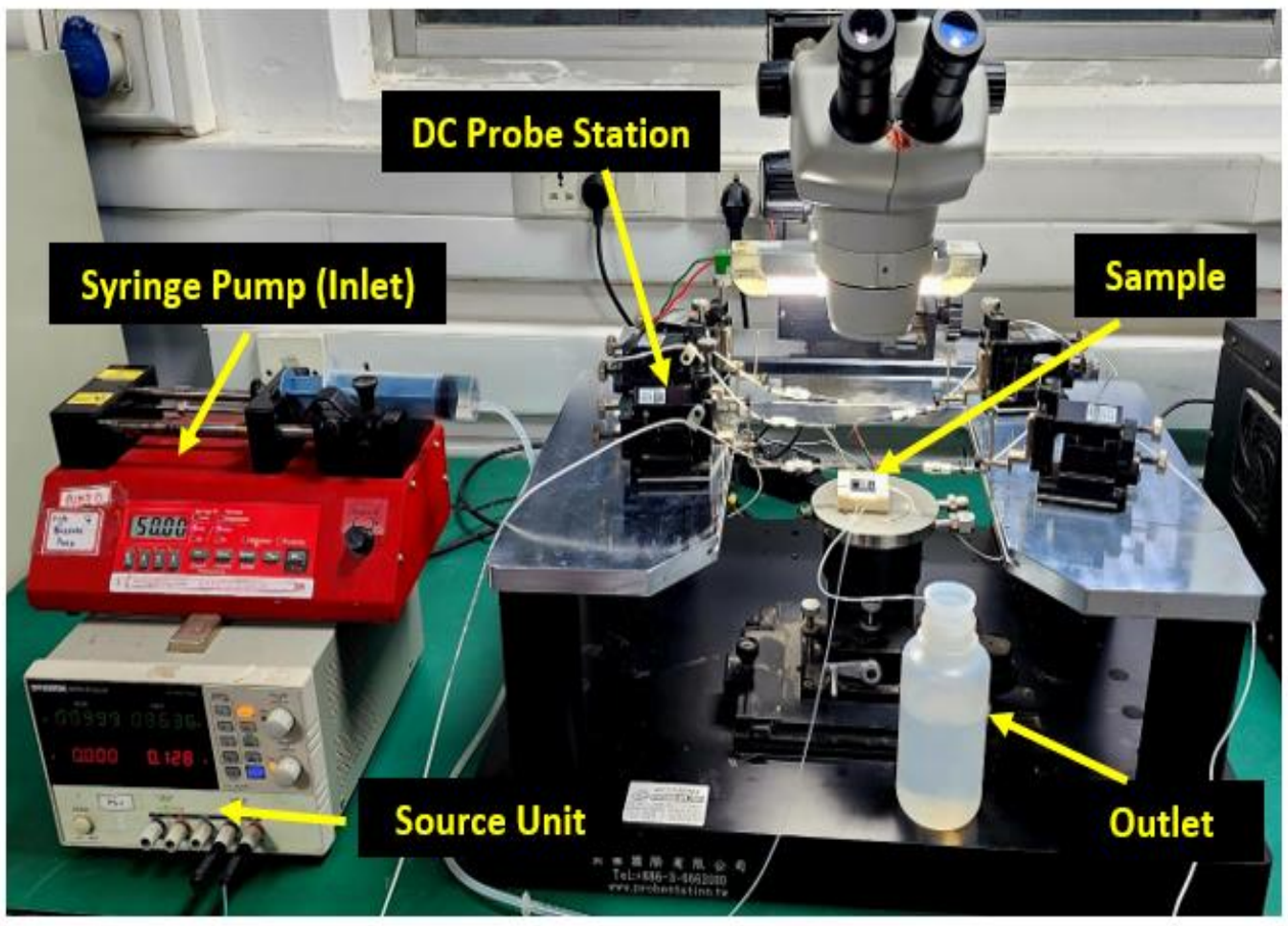

Fig. 3.

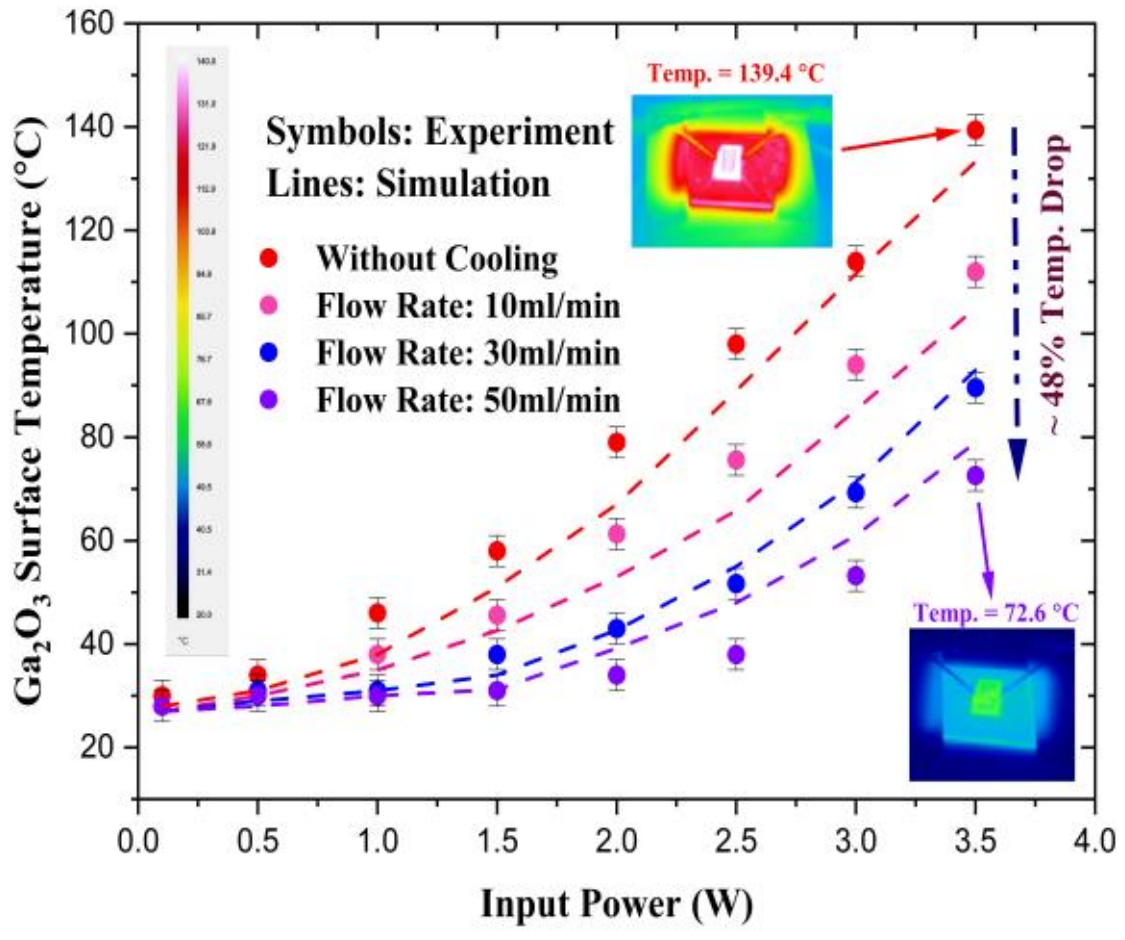
Symbols: Experiment
Lines: Simulation
Without Cooling
Flow Rate: 10ml/min
Flow Rate: 30ml/min
Flow Rate: 50ml/min
Temp. = 139.4 °C
Temp. = 72.6 °C
~ 48% Temp. Drop
$Ga_2O_3$ Surface Temperature (°C)
Input Power (W)

Fig. 4.

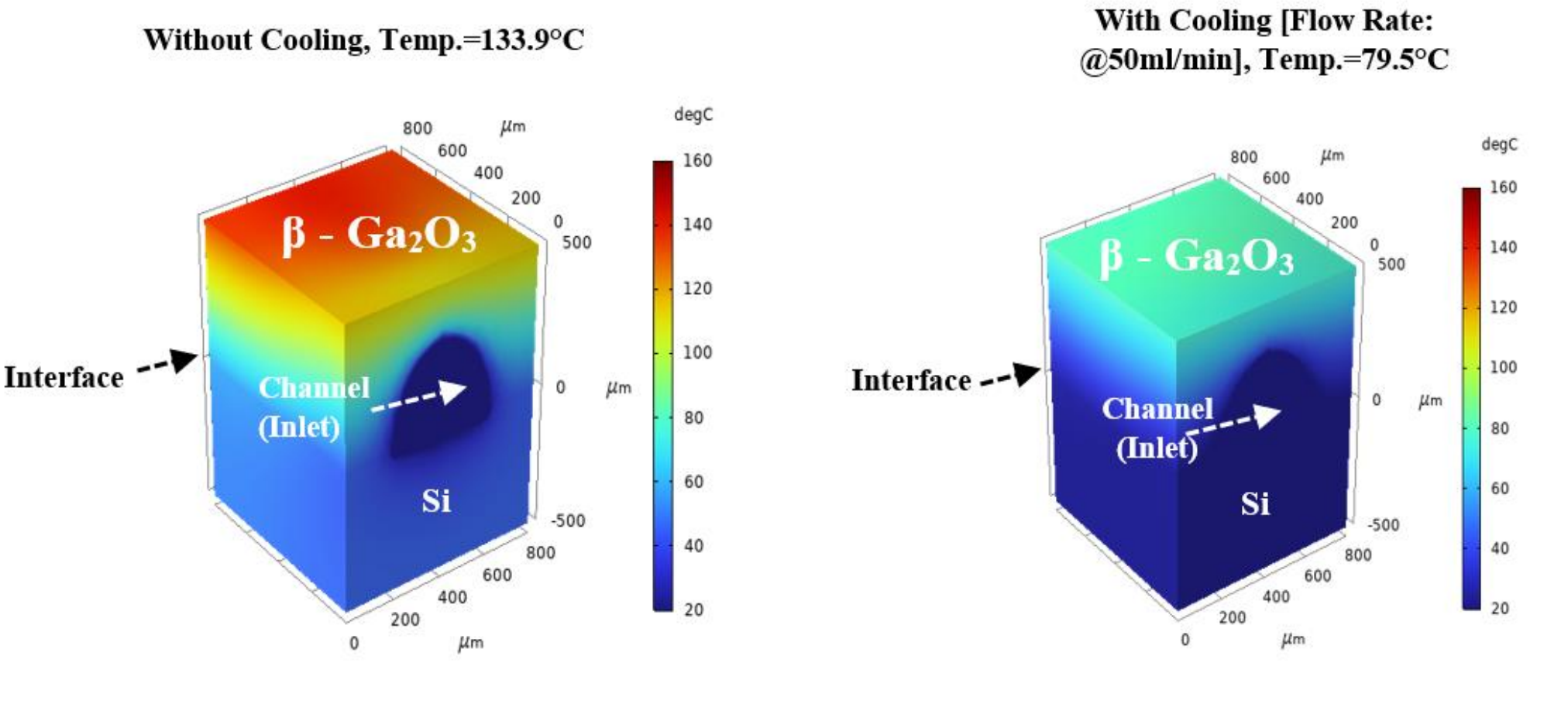

(a) (b)

Fig. 5.